\documentclass[
    ,final            
  ,numberedheadings 
  ]
  {aipproc}

\layoutstyle{6x9}

\usepackage{graphicx}

\def\ET{\mbox{$E_T$}}
\def\pT{\mbox{$p_T$}}

\def\sqrtsNN{\mbox{$\sqrt{s_\mathrm{NN}}$}}

\def\Nbinary{\mbox{$\mathrm{N}_\mathrm{bin}$}}
\def\NbinaryMean{\mbox{$\langle\Nbinary\rangle$}}

\def\kzeros{\mbox{$K^0_s$}}

\def\lam{\mbox{$\Lambda$}}
\def\lambar{\mbox{$\bar\Lambda$}}

\def\vtwo{\mbox{$v_2$}}

\def\RAB{\mbox{$R_{AB}(\pT)$}}
\def\RCP{\mbox{$R_{CP}(\pT)$}}

\def\TAB{\mbox{$T_{AB}$}}

\def\sigmappinel{\mbox{$\sigma^{pp}_{inel}$}}

\def\lt{\mbox{$<$}}
\def\gt{\mbox{$>$}}

\begin{document}

\title{Jets and high \pT\ hadrons in dense matter: recent results from STAR}

\author{Peter Jacobs}{
  address={Lawrence Berkeley National Laboratory, 1 Cyclotron Road, Berkeley CA 94720}
  ,altaddress={for the STAR Collaboration} 
}

\author{Jennifer Klay}{
  address={Lawrence Berkeley National Laboratory, 1 Cyclotron Road, Berkeley CA 94720}
  ,altaddress={for the STAR Collaboration} 
}


\begin{abstract}
We review recent measurements of high transverse momentum (high \pT)
hadron production in nuclear collisions by the STAR Collaboration at
RHIC. The previously observed suppression in central Au+Au collisions
has been extended to much higher \pT. New measurements from d+Au
collisions are presented which help disentangle the mechanisms
responsible for the suppression. Inclusive single hadron spectra are
enhanced in d+Au relative to p+p, while two-particle azimuthal
distributions are observed to be similar in p+p, d+Au and peripheral
Au+Au collisions. The large suppression of inclusive hadron production
and absence of the away-side jet-like correlations in central Au+Au
collisions are shown to be due to interactions of the jets with the
very dense medium produced in these collisions.
\end{abstract}

\maketitle


\section{Introduction}

High energy partons propagating through matter are predicted to lose
energy via induced gluon radiation, with the magnitude of the
energy loss depending linearly on the color charge density of the
matter\cite{EnergyLoss}. This phenomenon may provide a sensitive probe
of the medium generated in high energy heavy ion collisions, where a
Quark-Gluon Plasma (QGP) is expected to form if sufficiently high
energy density is achieved. The high energy partons in such collisions
result from the hard scattering of quarks or gluons from the incoming
nuclei and are observed experimentally as correlated ``jets'' of
hadrons having large transverse momentum (\pT) with respect to the
beam direction. The experimental challenge is to measure partonic
energy loss (``jet quenching'') in the extremely complex environment
of high energy nuclear collisions.

In these talks we discuss recent progress towards the measurement of jet
quenching in high energy nuclear collisions by the STAR experiment at
the Relativistic Heavy Ion Collider (RHIC) at Brookhaven National
Laboratory. At the values of jet \ET\ accessible at RHIC energies with
the currently achieved integrated luminosities, full jet
reconstruction with good energy resolution is difficult. We therefore
utilize known features of jet fragmentation to study jet quenching, in
particular the inclusive spectrum of high \pT\ (``leading'') hadrons
and the angular correlation of pairs of high \pT\ hadrons. We compare
their distributions in centrality-selected Au+Au collisions to those
in d+Au and non-singly diffractive (NSD) p+p collisions, all at
\sqrtsNN=200 GeV. Unless otherwise specified, the results reported here are for
unidentified charged hadrons, measured in the large cylindrical STAR
Time Projection Chamber with a 0.5T solenoidal magnetic
field\cite{STARNIM}. Related results from STAR are discussed in
\cite{Miller,Tang}.


\section{Suppression of inclusive charged hadrons}

\begin{figure}
\centering
\includegraphics[width=.5\textwidth]{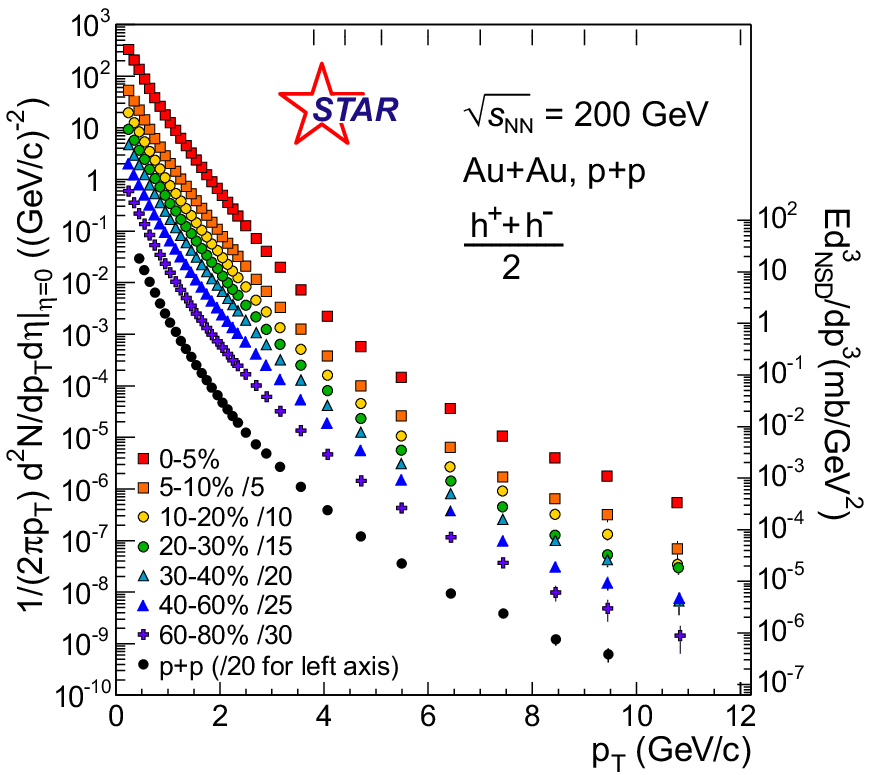}
\hspace{.05\textwidth}
\includegraphics[width=.45\textwidth]{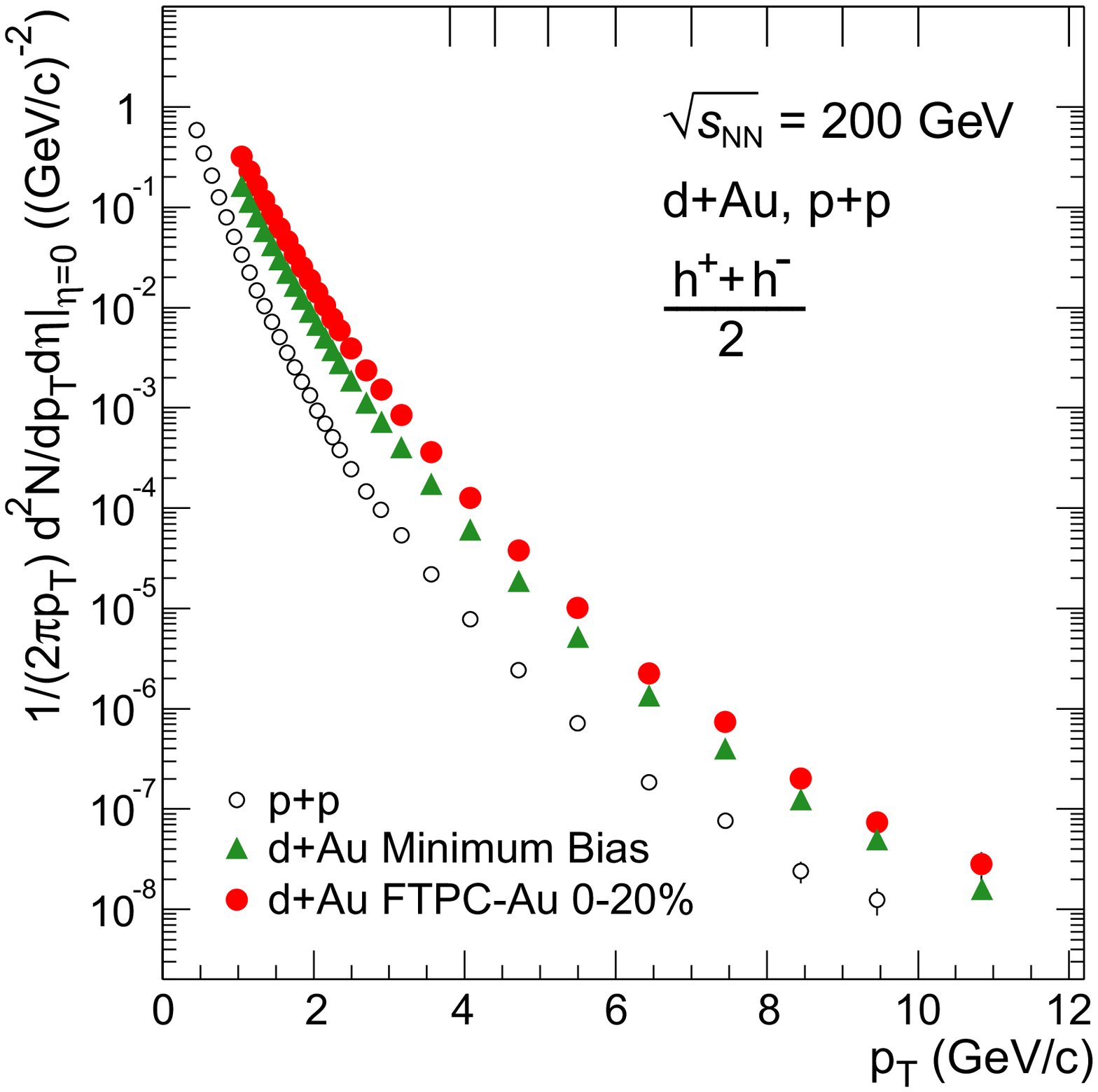}
\caption{
Invariant inclusive \pT\ distributions of charged hadrons at
\sqrtsNN=200 GeV. Left: NSD p+p and centrality-selected Au+Au
collisions\cite{STARHighpt200}. Right: central and minimum bias d+Au collisions, and same
p+p spectrum as left panel\cite{STARdAu}.}
\label{Spectra}
\end{figure}

Jets occasionally fragment with a single hadron carrying a large
fraction of the total jet energy. The resulting inclusive hadron
distributions exhibit power-law ($1/\pT^n$) shapes characteristic of
the underlying perturbative QCD (pQCD) processes. Figure 
\ref{Spectra}
shows the event-normalized invariant inclusive charged hadron
distributions as a function of \pT\ for NSD p+p and
centrality-selected Au+Au and d+Au
collisions\cite{STARHighpt200,STARdAu}. The centrality bins correspond
to the indicated percentiles of the total cross section: 0-5\% for
Au+Au indicates the most central and 60-80\% the most peripheral
collisions. Power-law shapes are clearly observed in all cases.

Nuclear effects on hadron production in d+Au and Au+Au collisions are measured
through comparison to the p+p reference spectrum using the ratio
\begin{equation}
\label{eqRAB}
\RAB=\frac{d^2N/d{\pT}d\eta}{\TAB\,{d}^2\sigma^{pp}/d{\pT}d{\eta}}\ ,
\end{equation}
\noindent
where $d^2N/d{\pT}d\eta$ is the differential yield per event in the
nuclear collision $A+B$, \TAB=\NbinaryMean/\sigmappinel\ describes the
nuclear geometry, and ${d}^2\sigma^{pp}/d{\pT}d{\eta}$ for p+p
inelastic collisions is determined from the p+p
measurement. \NbinaryMean\ is the mean number of binary NN
interactions for the given centrality class of $A+B$ collisions. In the
absence of nuclear effects such as shadowing, the Cronin effect, or
gluon saturation, hard process rates are expected to scale with
\NbinaryMean, and \RAB=1.

\begin{figure}
\centering
\includegraphics[width=.45\textwidth]{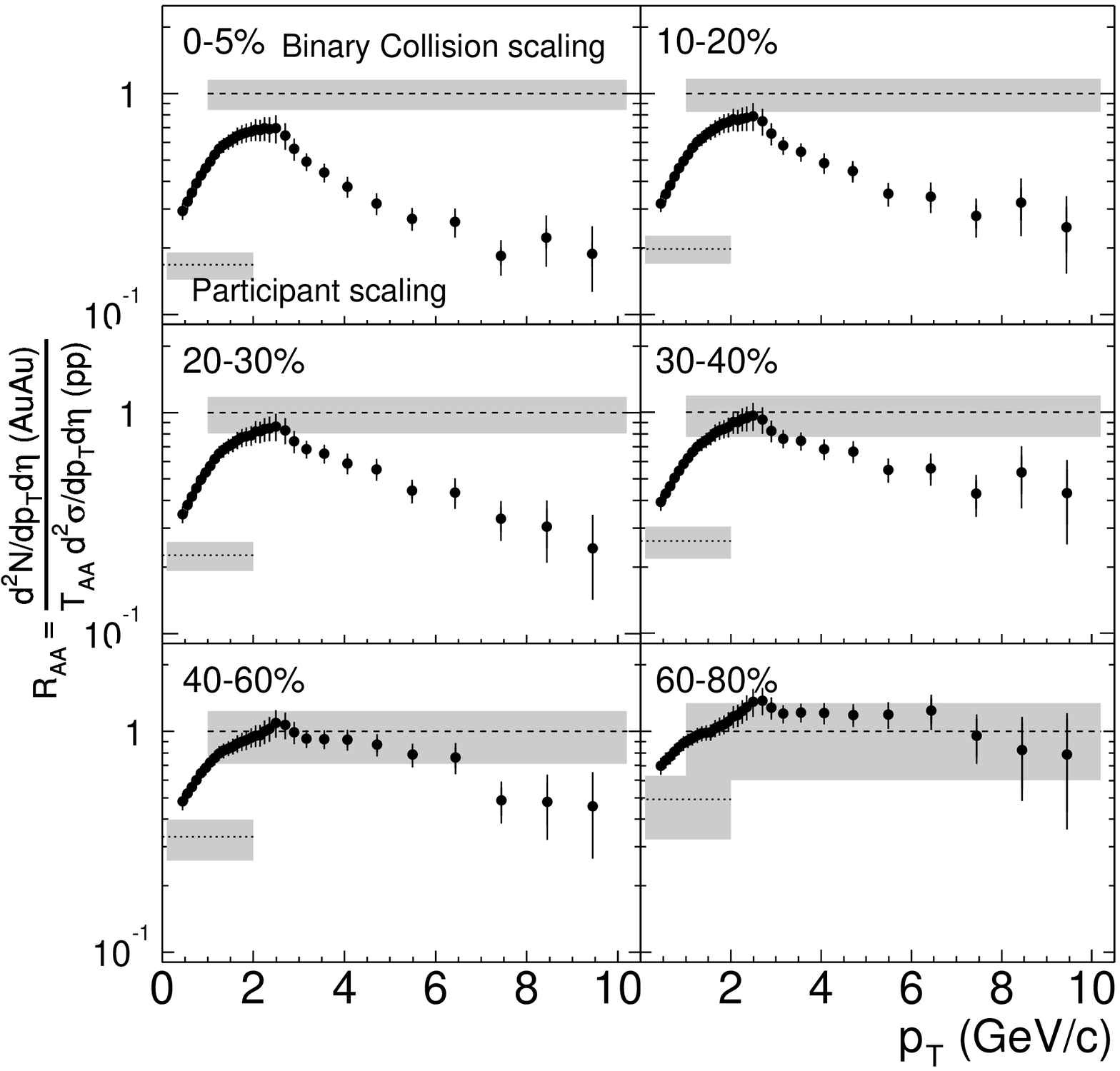}
\hspace{.05\textwidth}
\includegraphics[width=.5\textwidth]{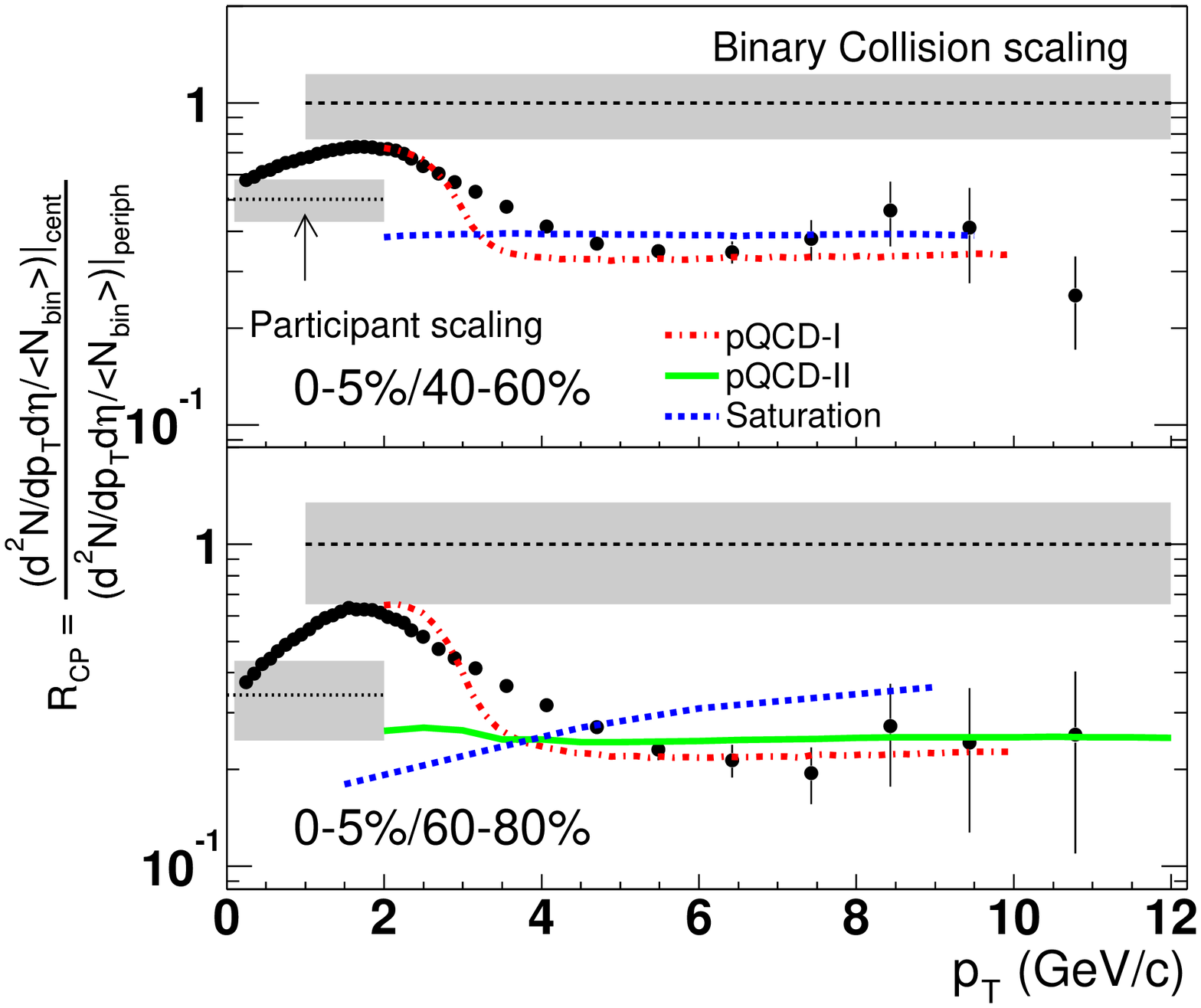}
\caption{Left: \RAB\ (eq. \ref{eqRAB}) for Au+Au relative to p+p collisions\cite{STARHighpt200}. 
Right: \RCP\ from Au+Au collisions\cite{STARHighpt200}.}
\label{RAARCP}
\end{figure}

Figure \ref{RAARCP}, left panel, shows \RAB\ for centrality-selected
Au+Au relative to p+p collisions. The error bars indicate the statistical and
systematic uncertainties of the spectra, while the bands indicate the
uncertainty due to the geometrical scaling factor \TAB. At the highest
\pT, hadron suppression of approximately a factor 5 is observed for
the most central collisions: inclusive hadron production is strongly
suppressed at high \pT\ in central Au+Au collisions. For the most
peripheral collisions \RAB\ is consistent with unity, while
intermediate centralities interpolate smoothly between the extremes.

Figure \ref{RAARCP}, right panel, shows the related quantity \RCP, the
\NbinaryMean-scaled ratio of particle yields in central relative to
peripheral collisions, which also exhibits large suppression of
inclusive hadron production in central collisions. Figure \ref{RAARCP}
also shows the results of three theoretical calculations: two models
based on pQCD incorporating shadowing, the Cronin effect and jet
quenching in dense matter (pQCD-I\cite{pQCDI}, pQCD-II\cite{pQCDII}),
and a model incorporating gluon saturation in the incoming Au
nuclei\cite{KLM}. The pQCD-based calculations contain one free
parameter, the energy density for central collisions, which is fit to
the data, yielding an initial density 30-50 times that of cold nuclear
matter \cite{pQCDI,pQCDII}. These calculations then successfully
describe the \pT\ and centrality dependence of the inclusive
suppression for \pT\gt5 GeV/c. The saturation model also describes the
suppression and its \pT-dependence for the 0-5\%/40-60\% ratio for
\pT\gt5 GeV/c. These precision STAR data on inclusive charged hadron
suppression, covering wide centrality and kinematic range, are
well-described by widely differing models which attribute the
suppression either to jet quenching in dense matter in the final state
or to gluon saturation in the initial state.

\begin{figure}
\centering
\includegraphics[width=.5\textwidth]{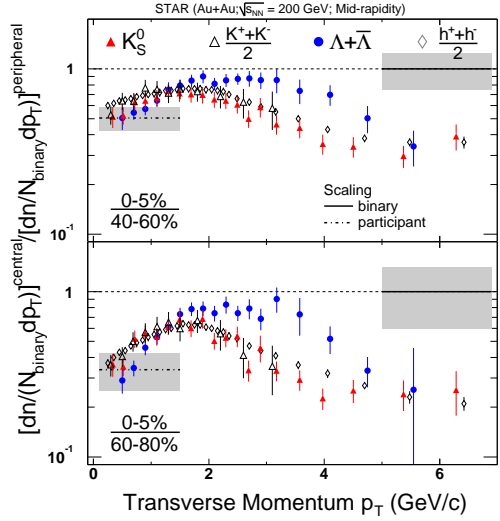}
\caption{\RCP\ from Au+Au collisions for \kzeros\ and \lam+\lambar \cite{STARStrangeRCP}, and for the charged hadron data 
also shown in Fig. \ref{RAARCP}, right panel.}
\label{KsLamRCP}
\end{figure}

In order to elucidate the particle-species dependence of the
suppression, Fig. \ref{KsLamRCP} shows \RCP\ for \kzeros\ and
\lam+\lambar\\cite{STARStrangeRCP} compared to the charged hadron \RCP\ already 
shown in Fig. \ref{RAARCP}. The mesons scale approximately as the charged
hadrons throughout, while the baryons exhibit a pronounced enhancement
in the region 2\lt\pT\lt4 GeV/c. The origin of this behavior is at
present not understood, though speculations include the Cronin effect
and non-perturbative mechanisms such as baryon junctions.  At \pT=5
GeV/c, all particle species are strongly suppressed by the same amount
in central collisions.

\section{Correlations}

\begin{figure}
\centering
\includegraphics[width=.45\textwidth]{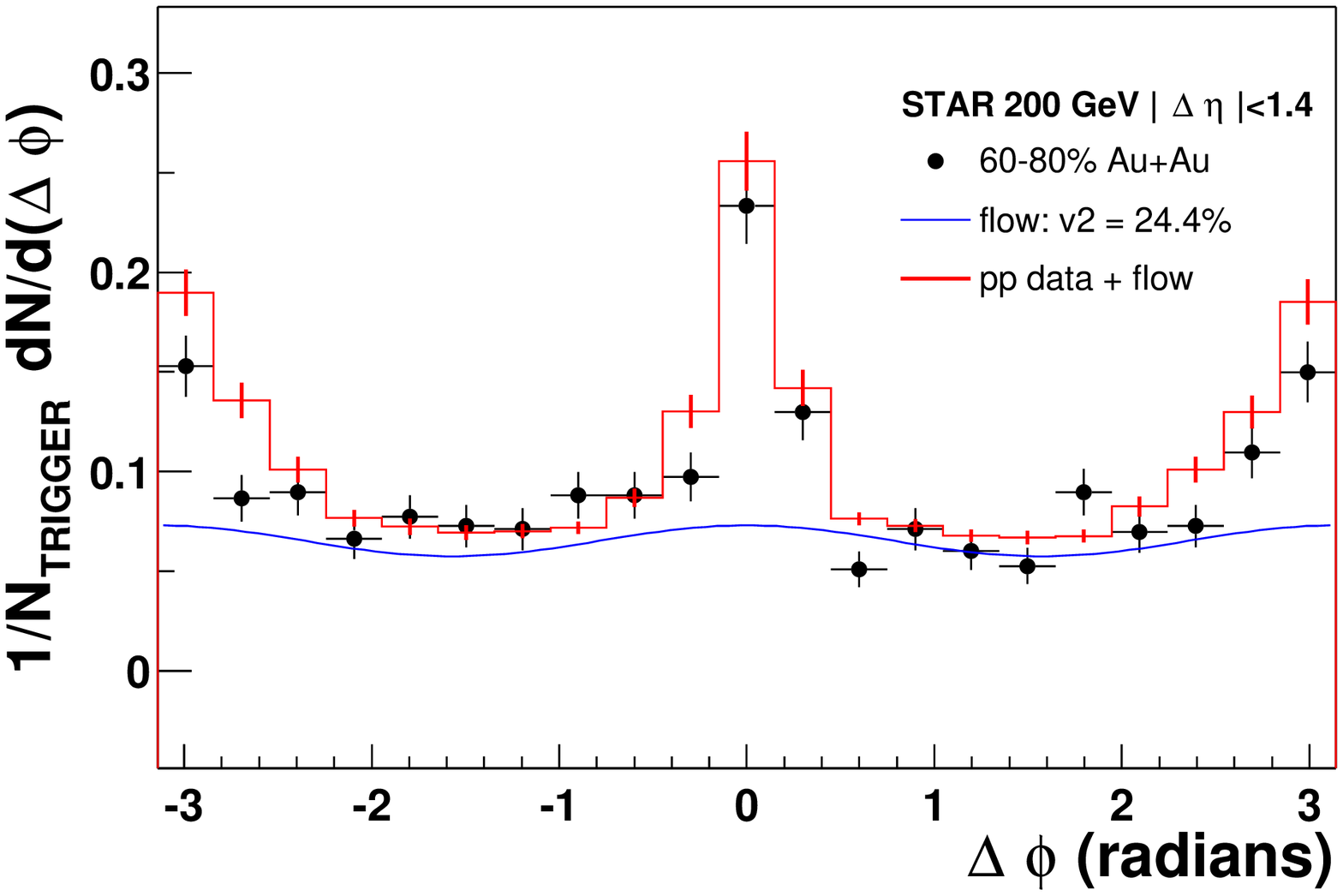}
\hspace{.1\textwidth}
\includegraphics[width=.45\textwidth]{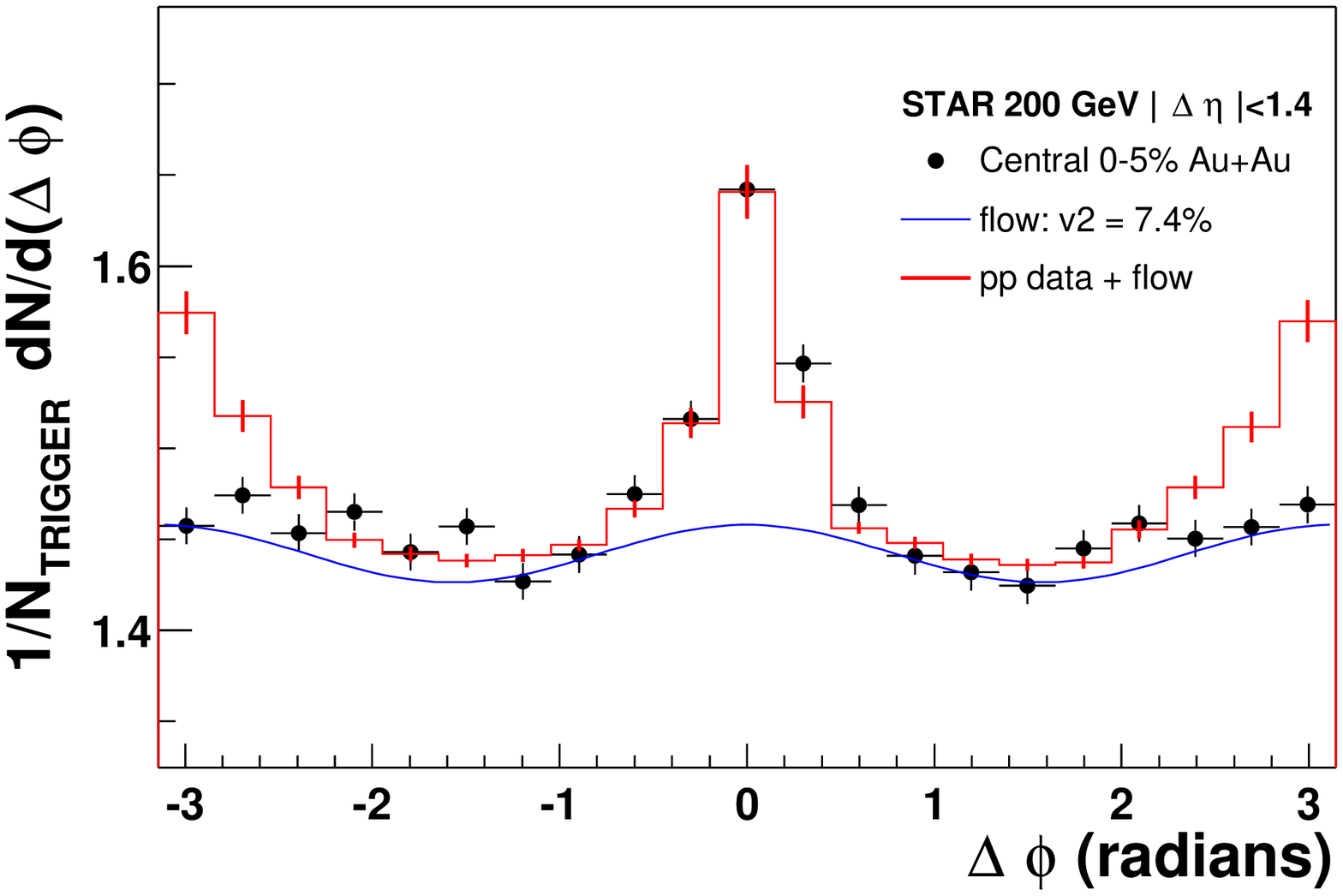}
\caption{Azimuthal correlations for peripheral (left) and central (right) 
Au+Au collisions compared to the pedestal and flow-scaled correlations
in p+p collisions\cite{STARHighptCorrelations}.}
\label{C2AuAu}
\end{figure}

The angular correlations of pairs of high-\pT\ charged particles can be used 
to study jets in the complex environment of heavy ion collisions\cite{Miller}.
Figure \ref{C2AuAu} shows the two-particle azimuthal distribution
$D(\Delta \phi)$, defined as
\begin{equation} 
D(\Delta \phi) \equiv \frac{1}{N_{trigger}}\frac{1}{\epsilon}
\frac{dN}{d(\Delta \phi)} ,
\end{equation} 
for peripheral (left panel) and central (right panel) Au+Au collisions 
and for p+p collisions (both panels).
Only particles within $|\eta|\lt0.7$ are included in the analysis. 
$N_{trigger}$ is the number of particles within 4\lt\pT(trig)\lt6 GeV/c, 
referred to as trigger particles. The distribution results from the 
correlation of each trigger particle with all associated particles in the 
same event having $2<p_T<p_T$(trig), where $\epsilon$ is the tracking 
efficiency of the associated particles. The normalization uncertainties are 
less than 5\%.  

In order to compare correlations in Au+Au with p+p, the p+p correlations 
are scaled up to the same pedestal value at $|\Delta \phi| \sim 
\pi/2$ as Au+Au and
superposed with a $\cos(2\Delta\phi)$ term that characterizes the
azimuthal anisotropy in non-central Au+Au collisions (elliptic flow
\cite{Tang}). The magnitude of this additional correlation is
measured independently and is given by the second coefficient \vtwo\
of a Fourier expansion of the azimuthal distribution relative to the
orientation of the reaction plane of the event, for particles having
\pT\lt2 GeV/c \cite{STARv2}. Figure \ref{C2AuAu} shows that the correlation 
strength at small relative angle ($\Delta \phi \sim 0$) in peripheral and central
Au+Au and at large relative angle ($\Delta \phi \sim \pi$) in
peripheral Au+Au are very similar to the scaled correlations in p+p
collisions.

The near-side peaks ($\Delta \phi \sim 0$) in all three collision systems 
are characteristic of jet fragmentation.  The away-side peak ($|\Delta \phi| 
\sim \pi$) from the back-to-back partner jet, apparent in p+p and peripheral 
Au+Au collisions is strongly suppressed in central Au+Au
collisions. These observations, together with the strong suppression
of inclusive production and large elliptic flow at high \pT\
\cite{Tang}, suggest a picture in which jets traversing the bulk of the medium 
produced in Au+Au collisions are absorbed (strong jet quenching), and
the observed jets are biased towards those generated on the surface and
heading outwards. However, other explanations, in particular the gluon
saturation picture, may also account for some or all of the observed
phenomena. Experimentally, these very different scenarios can be
discriminated through measurements of d+Au collisions, where initial
state effects in the Au nucleus remain but no final state dense medium
is generated.

\section{d+Au collisions}

pQCD-based models predict an enhancement in the production of high-\pT charged 
hadrons in d+Au collisions relative to binary-scaled p+p
collisions and little change in the back-to-back correlation
strength\cite{VitevdAu,WangdAu}, while one version of the saturation
model\cite{KharzeevdAu} predicts an inclusive suppression in d+Au of
about 30\% and possibly also suppression of the back-to-back strength
due to a mono-jet contribution \cite{BtoBsaturation}.

\begin{figure}
\centering
\includegraphics[width=.5\textwidth]{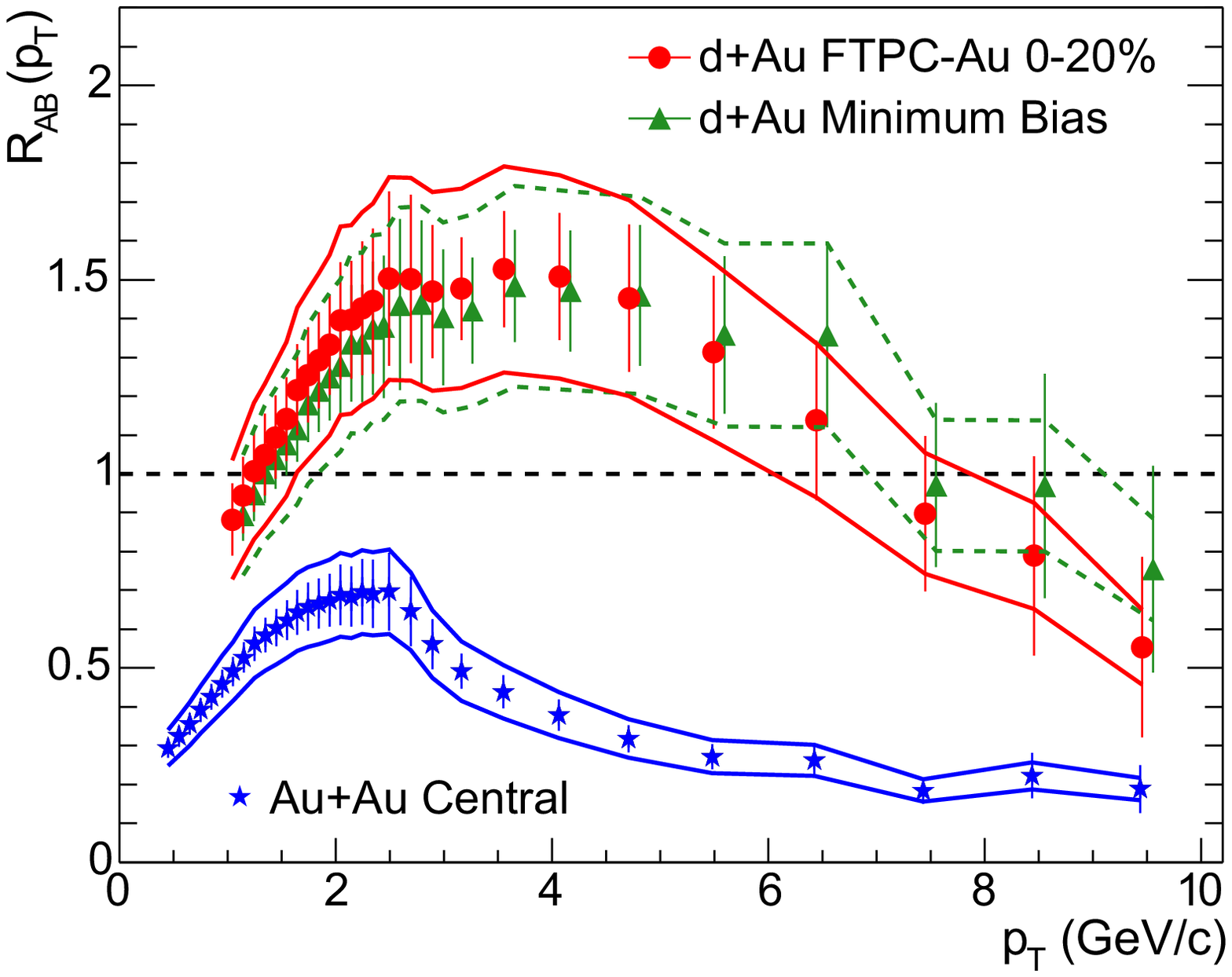}
\hspace{.05\textwidth}
\includegraphics[width=.45\textwidth]{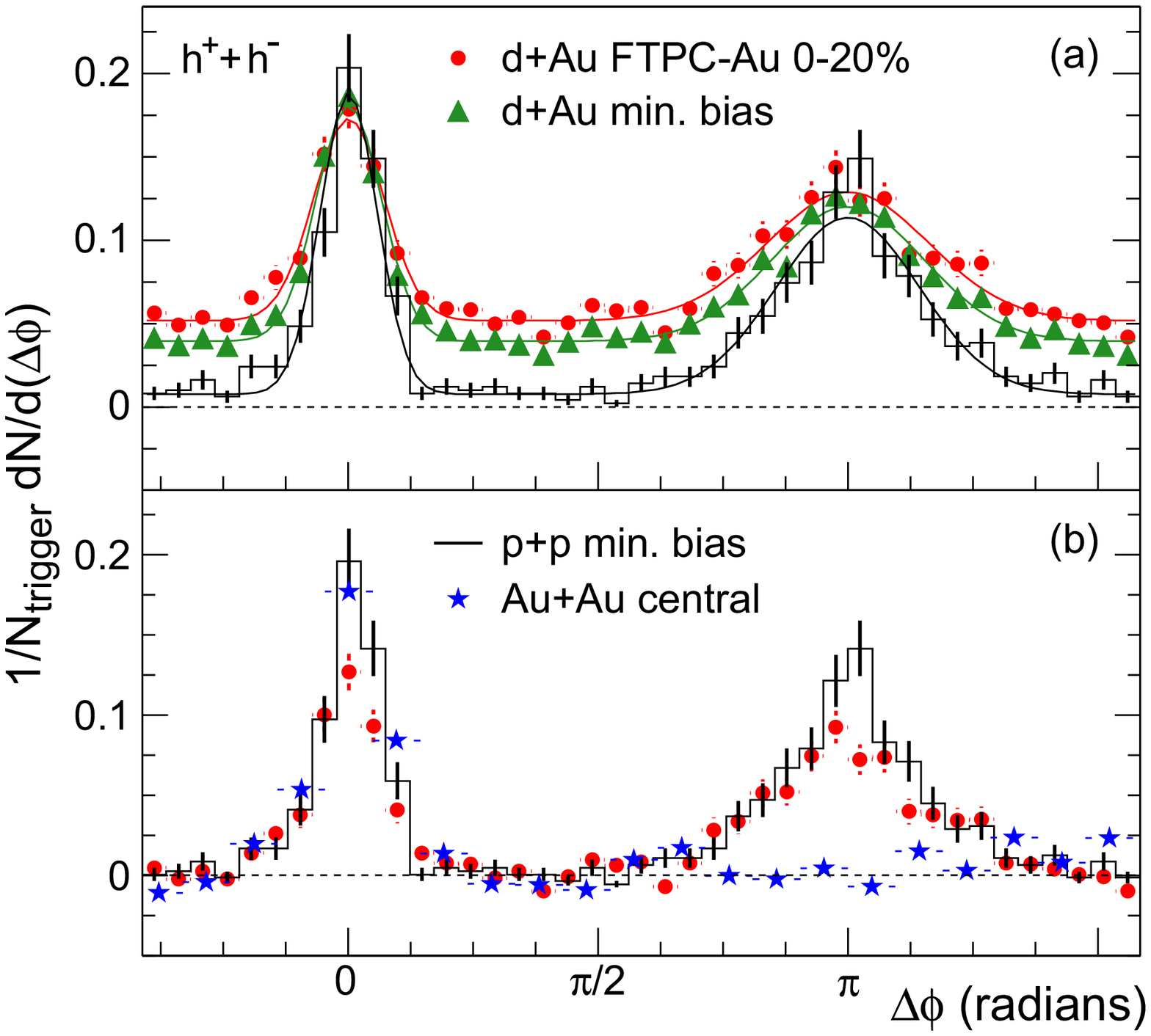}
\caption{\RAB\ and two-particle azimuthal distributions in d+Au 
collisions\cite{STARdAu}. The horizontal axes of the right panel are shifted by
$\pi/2$ relative to Fig. \ref{C2AuAu}.  }
\label{dAu}
\end{figure}

Figure \ref{dAu} shows \RAB (left) and the pedestal subtracted
two-particle azimuthal distributions (right) measured by STAR in
minimum bias and central d+Au collisions\cite{STARdAu}. \RAB\ exceeds
unity for 2\lt\pT\lt7 GeV/c, consistent with expectations from the
Cronin effect. However, no additional enhancement over p+p is observed
for central relative to minimum bias d+Au collisions. In the top right
panel, the azimuthal distributions are characterized by a fit to the
sum of near-side and back-to-back Gaussian peaks plus a constant. The
only significant difference between the p+p and d+Au correlations is
the growth in the pedestal value \cite{Miller}. The lower right panel shows the
pedestal- and flow-subtracted azimuthal distribution from central
Au+Au collisions along with the pedestal-subtracted distributions from
central d+Au and p+p.  The near-side peak is similar in all three
collision systems, while the away-side peak is suppressed only in
central Au+Au collisions.

\section{Summary}

The strong suppression at high \pT\ of the inclusive hadron yield and
back-to-back correlations in central Au+Au collisions at \sqrtsNN=200
GeV are not observed in d+Au collisions. The inclusive yield in d+Au
collisions is enhanced relative to binary-scaled p+p, consistent with
expectations from the Cronin effect, and the back-to-back correlations
show little variation relative to p+p. These results demonstrate
conclusively that the striking suppression phenomena observed in central Au+Au
collisions are due to the interaction of high energy partons or their
fragmentation products in the dense medium created in such collisions.






\def\etal{\mbox{$\mathrm{\it et\ al.}$}}


\end{document}